\documentclass[aps,pra,twocolumn,groupedaddress,showpacs]{revtex4}

\usepackage{amssymb}   
\usepackage{amsmath}   
\usepackage{graphicx} 

\begin{document}

\preprint{MPQ, BRISBANE}

\title{Entanglement renormalization and topological order}

\author{Miguel Aguado}
\affiliation{Max-Planck-Institut f\"ur Quantenoptik.
  Hans-Kopfermann-Str.~1.  D-85748 Garching, Germany}

\author{Guifr\'e Vidal}
\affiliation{School of Physical Sciences. The University of
  Queensland. Brisbane, QLD, 4072, Australia}

\date{\today}

\begin{abstract}
  The multi-scale entanglement renormalisation ansatz (MERA) is argued
  to provide a natural description for topological states of matter.
  The case of Kitaev's toric code is analyzed in detail and shown to
  possess a remarkably simple MERA description leading to distillation
  of the topological degrees of freedom at the top of the tensor
  network.  Kitaev states on an infinite lattice are also shown to be
  a fixed point of the RG flow associated with entanglement
  renormalization.  All these results generalize to arbitrary quantum
  double models.
\end{abstract}


\maketitle

Renormalization group (RG) transformations aim to obtain an effective
description of the large distance behavior of extended systems
\cite{Fisher}. In the case of a system defined on a lattice, this can
be achieved by constructing a sequence of increasingly coarse-grained
lattices $\{ {\mathcal{L}}_0, \, {\mathcal{L}}_1, \, {\mathcal{L}}_2,
\, \cdots \}$, where a single site of lattice ${\mathcal{L}}_\tau$
effectively describes a block of an increasingly large number
$n_{\tau} \sim \exp(\tau)$ of sites in the original lattice
${\mathcal{L}}_0$ \cite{Kadanoff}.  Real-space RG methods can, in
particular, be applied to study quantum systems at zero temperature,
in which case each site of ${\mathcal{L}}_\tau$ is represented by a
Hilbert space ${\mathcal{K}}_{\tau}$ \cite{Wilson}.  There the goal is
to identify the local degrees of freedom relevant to the physics of
the ground state and to retain them in the Hilbert space
${\mathcal{K}}_\tau$, whose dimension $d_\tau$ must be large enough to
describe this physics.  A severe problem of such approach is that in
$D \geq 2$ dimensions, $d_\tau$ must grow (doubly) exponentially in
$\tau$ \cite{large_dimension} as a result of the accumulation of
short-range entanglement at the boundary of the block.

\begin{figure}[h]
  \includegraphics[width=8.5cm]{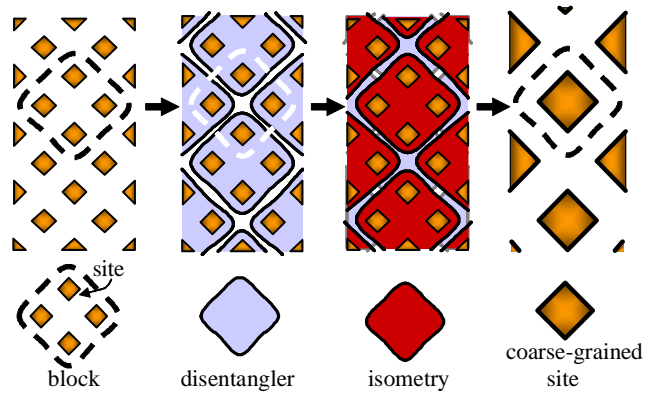}
  \caption{RG transformation based on entanglement renormalization. In
    order to build an effective site from a block of four sites, we
    first apply disentanglers between sites of the block and
    surrounding sites. In this way part of the short-ranged
    entanglement between the block and its surroundings is
    removed. Then we coarse-grain the four sites into one by means of
    an isometry that selects the subspace ${\cal K'} \subseteq {\cal
      K}^{\otimes 4}$ to be kept. We show the case of a tilted square
    lattice in preparation for the toric code where, in addition, each
    site will contain four qubits.}\label{figure:2DMERA}
\end{figure}

\emph{Entanglement renormalization} \cite{ER} is a novel real-space RG
transformation recently proposed in order to solve the above
difficulties. Its defining feature is the use of {\em disentanglers}
prior to the coarse-graining step. These are unitary operations,
acting on the interface of the blocks defined by the RG procedure,
that reduce the amount of entanglement in the system, see figure
\ref{figure:2DMERA}.  A major achievement of the approach is that, when
applied to a large class of ground states in both one \cite{ER} and
two \cite{2DMERA} spatial dimensions, the dimension $d_{\tau}$ is seen
not to grow with $\tau$. A steady $d_{\tau}$ is made possible by the
disentangling step and has several implications \cite{ER,2DMERA}.  It
means that, in principle, the resulting RG transformation can be
iterated indefinitely at a constant computational cost, allowing for
the exploration of arbitrarily large length scales. In addition, the
system can be compared with itself at different length scales, and
thus we can study RG flows in the space of ground state or Hamiltonian
couplings.  Finally, a constant $d_{\tau}$ also leads to an efficient
representation of the system's ground state in terms of a tensor
network, the {\em multi-scale entanglement renormalization ansatz}
(MERA) \cite{MERA}.

At zero temperature, strongly correlated quantum systems appear
organized in a plethora of {\em phases} or {\em orders}, including
{\em local symmetry breaking} orders and {\em topological} orders
\cite{WenBook}.  Local symmetry breaking phases are described by a
symmetry group and a local order parameter, and they are associated
with the physical mechanism of condensation of point-like objects.
Transitions between two such phases or orders involve a change in the
symmetry, as described by Landau's theory.  A simple picture emerges
from the perspective of entanglement renormalization \cite{ER,2DMERA}:
under successive iterations of the RG transformation, ground states
with local symmetry breaking order progressively lose their
entanglement and eventually converge to a trivial fixed point, namely
an unentangled ground state.  On the other hand, critical ground
states describing transitions between these phases are non-trivial
---that is, entangled--- fixed points of the RG transformation.  In
either case, the MERA provides an efficient, accurate representation
of the ground state.

Topological phases are fundamentally different from local symmetry
breaking phases \cite{WenBook}.  They do not stem from (the breakdown
of) local group symmetries, but their {\em topological order} is
linked to more complex mathematical objects, like tensor categories,
topological quantum field theory, and quantum groups.  Physically,
topological phases exhibit gapped ground levels with robust degeneracy
dependent only on the topology of the underlying space. This, and the
fact that excitations above the ground level possess anyonic
statistics, boosts the interest of these phases as scenarios for
topological quantum information storage and processing.  Condensation
of string-like objects (in the so-called string-net models, see
\cite{string-net}) has been proposed as a general mechanism
controlling topological phases.  As may be expected, such profound
differences are also reflected in the way the ground state is
entangled.  Specifically, the notion of topological entanglement
entropy \cite{kitaev-preskill} (the subleading term in a
large-perimeter expansion of the entanglement entropy of a system) has
arisen as a quantitative measure of the ground state entanglement due
to topological effects.  Systems with topological order thus provide
an unexplored scenario for entanglement renormalization techniques.

The purpose of this Letter is to establish entanglement
renormalization and the MERA as valid tools also for the description
and investigation of topological phases of matter.  For simplicity, we
analyze in detail Kitaev's toric code \cite{kitaev03}, a four-fold
degenerate ground state widely discussed in the context of quantum
computation and closely related to $\mathbb{Z}_2$ lattice gauge theory
\cite{wegner} and to the simplest of Levin-Wen's models for string-net
condensation \cite{string-net}. We show the following: ($i$) a MERA
with finite, constant $d_{\tau}$ can represent the toric code
\emph{exactly}; ($ii$) at each iteration of the RG transformation,
entanglement renormalization factors out local degrees of freedom from
the lattice, while leaving the topological degrees of freedom
untouched; $(iii)$ the MERA representation of the four ground states
is identical except in its top tensor, which stores the topological
degrees of freedom; and ($iv$) in an infinite system, the toric code
is the fixed point of this RG transformation. All these results also
hold for more complicated models, such as quantum double lattice
models, that we discuss in the appendix.  We conclude that the MERA is
naturally fitted to represent states with topological order, and the
entanglement renormalization offers a new, useful framework for
further studies.

Following \cite{kitaev03}, we consider a square lattice $\Lambda$ on
the torus, with spin-$1/2$ (qubit) degrees of freedom attached to each
link.  The Hamiltonian
\begin{equation}\label{kitaevhamiltonian}
	H = -\sum_+ A_+ - \sum_\square B_\square
\end{equation}
is a sum of constraint operators associated with vertices `$+$' and
plaquettes `$\square$,' namely
\begin{equation}\label{kitaevstabilizers}
  A_+
=
  \prod_{i \in +} X_i ,
\quad
  B_\square
=
  \prod_{i \in \square} Z_i .
\end{equation}
Stabilizers $A_+$ act as a simultaneous spin flip in all four qubits
adjacent to a given vertex.  Stabilizers $B_\square$ yield the product
of group assignments $\pm 1$ at the four qubits around a plaquette.
All stabilizers commute with each other and have eigenvalues $\pm 1$.
Hamiltonian (\ref{kitaevhamiltonian}) is gapped, and states in the
ground level (Kitaev states) are simultaneous eigenstates of all
$A_+$, $B_\square$ with eigenvalue $+1$.  The degeneracy of the ground
level (i.e., the number of Kitaev states) depends on the topology of
the manifold underlying the lattice. If this manifold is a
topologically nontrivial Riemann surface, information is encoded in
nontrivial cycles, since operators $\prod_{ i \in {\mathcal{C}}_{a,b}
} Z_i$, where ${\mathcal{C}}_{a,b}$ are nontrivial cycles along bonds
of the lattice, commute with all stabilizers.  Besides, such operators
along homologically equivalent nontrivial cycles ${\mathcal{C}}_a$,
${\tilde{\mathcal{C}}}_a$ have the same action on Kitaev states.
Hence, for a torus, two logical qubits are encoded in the action of
these operators.

\begin{figure}[h]
\centering
\scalebox{.5}{%
\includegraphics{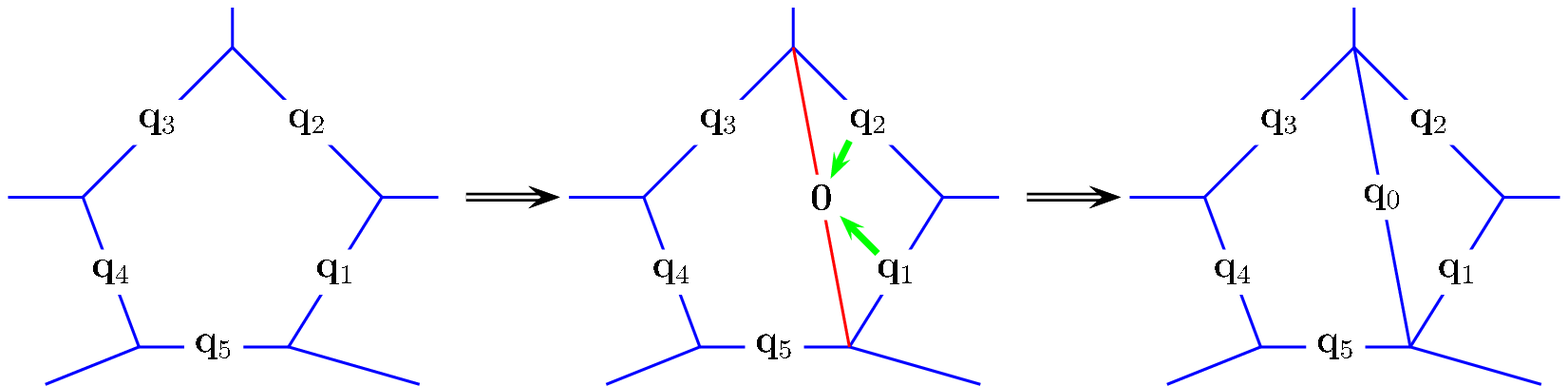}
}

\vspace{1.2cm}

\scalebox{.5}{%
\includegraphics{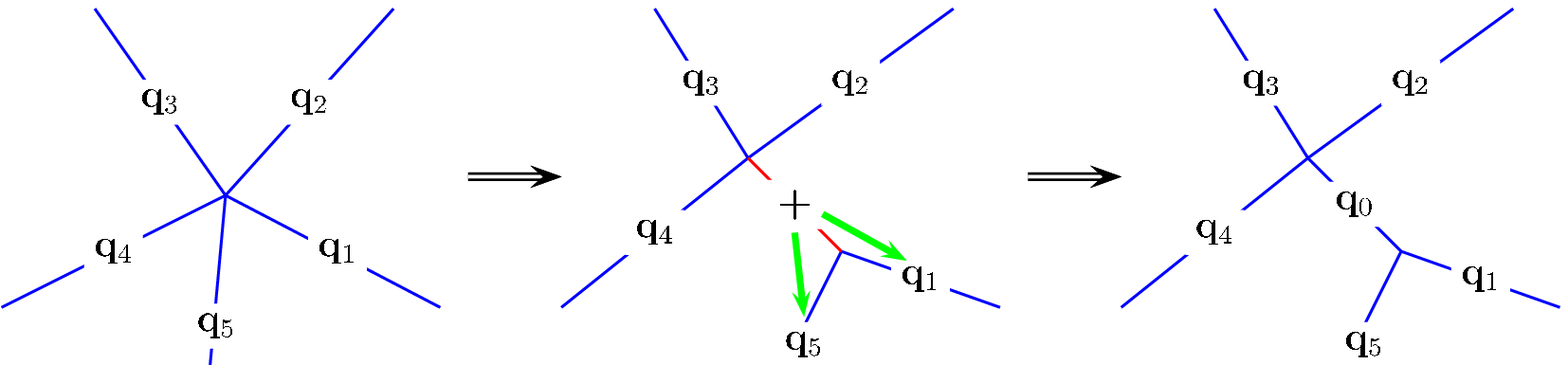}
}
\caption{\label{figure:elemmoves} %
  Elementary moves adding plaquettes and vertices to a toric code.
  Arrows stand for CNOT operations.}
\end{figure}

Kitaev states are efficiently written in terms of their stabilizers.
The stabilizer formalism \cite{gottesman} also provides us with a
useful language to analyse the action of operators on Kitaev states,
and has proved instrumental in finding an exact MERA.  The key
observation to this purpose is that there exist `elementary moves'
\cite{dennis}, minimal deformations of the lattice and its Kitaev
states, that respect the topological characteristics of the
code. These moves consists of addition or removal of faces and
vertices together with qubits, and can be written in terms of
controlled-NOT (CNOT) operators, whose adjoint action has a very
simple expression in terms of stabilizers:
\begin{align}\label{stabilizercnot}
  I \otimes Z
&\leftrightarrow
  Z \otimes Z ,
\quad
  Z \otimes I
\mapsto
  Z \otimes I , \\
  I \otimes X
&\mapsto
  I \otimes X ,
\quad
  X \otimes I
\leftrightarrow
  X \otimes X .
\end{align}

Figure \ref{figure:elemmoves} depicts the construction of elementary
moves. The creation of a face is achieved by introducing a new spin in
a plaquette.  Arrows stand for CNOT operators from control qubits (all
qubits in one of the semiplaquettes) to the target qubit (the new
qubit, introduced in state $\lvert 0 \rangle$).  The following
transformation of stabilizers holds (the new site is denoted as $n$):
\begin{align}\label{addplaquettestabilizers}
  Z_1 Z_2 Z_3 Z_4 Z_5 
&\longmapsto
  Z_1 Z_2 Z_3 Z_4 Z_5  , \\
  Z_n
&\longmapsto
  Z_1 Z_2 Z_n ,
\end{align}
which ensures plaquette constraints are obeyed. Similarly, the two
relevant vertex constraints are extended to the new qubit. The
creation of a new vertex is achieved instead by introducing a new
qubit in state $\lvert + \rangle$. This qubit now plays the role of
control for CNOTs acting on the qubits adjacent to one of the split
vertices. Stabilizers transform as
\begin{align}\label{addvertexstabilizers}
  X_1 X_2 X_3 X_4 X_5
&\longmapsto
  X_1 X_2 X_3 X_4 X_5 , \\
  X_n
&\longmapsto
  X_5 X_1 X_n ,
\end{align}
which is again compatible with the code constraints. Both final sets
of operators are the correct stabilizers for the code in the modified
lattice (remember that $X^2 = Z^2 = I$.) Similarly, the two relevant
paquette constraints are extended to the new qubit.
 
\begin{figure}
\centering
(a)%
\scalebox{.8}{%
%
%
\includegraphics{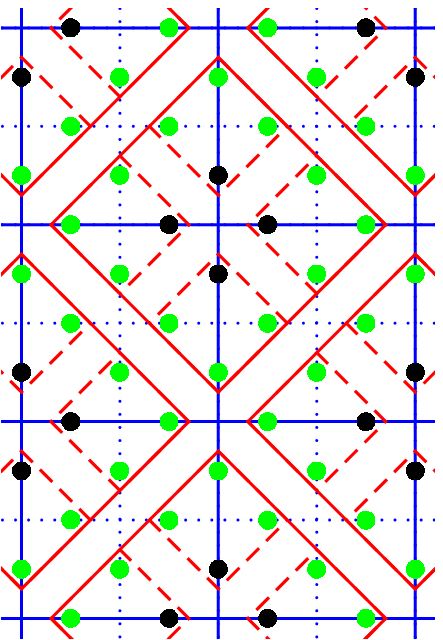}
}%
\hspace{.5cm}%
(b)%
\scalebox{.8}{%
%
%
\includegraphics{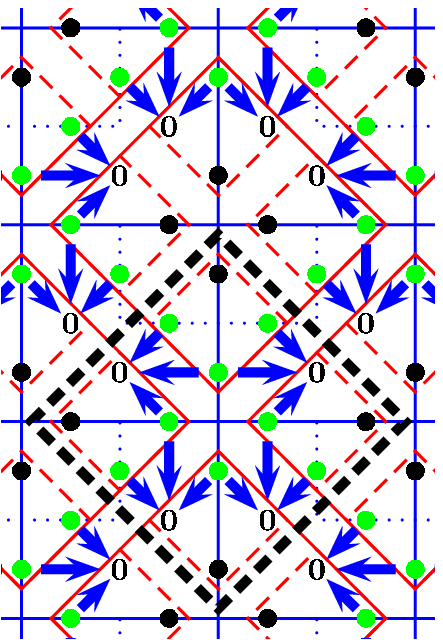}
}
\caption{\label{figure:meraa}%
  (a) The square lattice $\Lambda$ for the toric code, with qubits
  (dots) on the links, is reorganized into a tilted square lattice
  ${\mathcal{L}}_0$ where each site is made of four qubits. The
  lattice constant is doubled (dotted lines dissapear) after the RG
  transformation, which produces a new four-qubit site for lattice
  ${\mathcal{L}}_1$ from every block of sixteen qubits (the twelve
  light qubits in the block are decoupled in known product states).
  (b) First step of the RG transformation: Disentanglers.  Arrows
  stand for simultaneous CNOT operators from control to target qubits.
  Disentanglers act on sixteen-qubit domains overlapping with four
  blocks each (thick dashed line, cf.~figure \ref{figure:2DMERA}.)
  Four qubits per block decouple in state $\lvert 0 \rangle$.}
\end{figure}

\begin{figure}
\centering
%
%
(a)%
\scalebox{.8}{%
%
\includegraphics{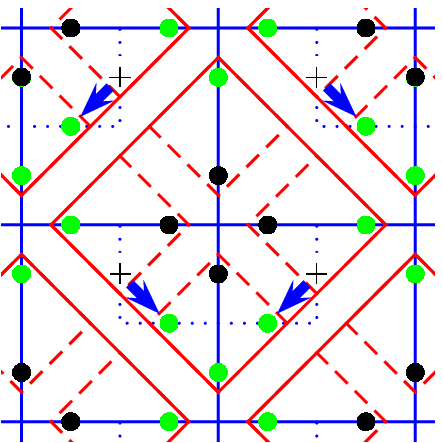}
}%
(b)%
\scalebox{.8}{
%
\includegraphics{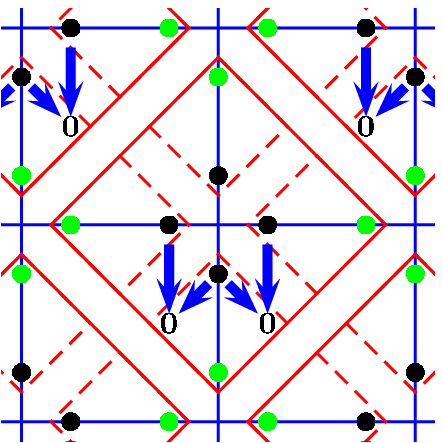}
}

\vspace{.3cm}

(c)%
\scalebox{.8}{
%
\includegraphics{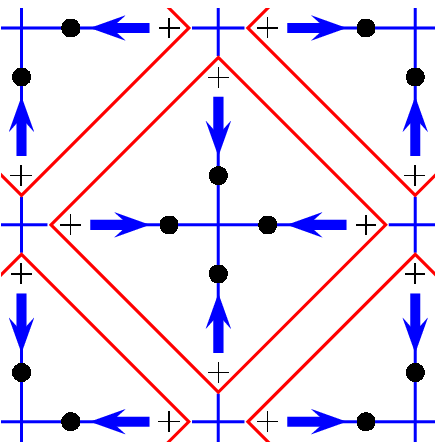}
}%
(d)%
\scalebox{.8}{
%
\includegraphics{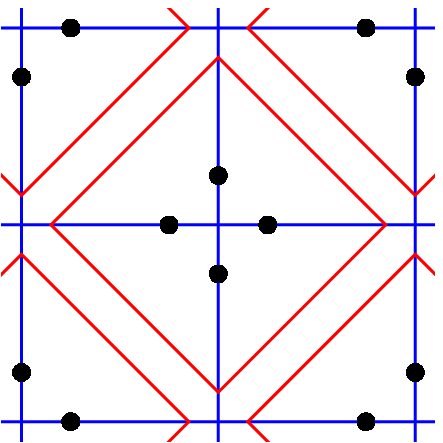}
}
\caption{\label{figure:merab}%
  (a)--(c) Second step of the RG transformation: Isometries.  (a) Two
  qubits per block decouple in state $\lvert + \rangle$. (b) Two more
  qubits per block decouple in state $\lvert 0 \rangle$. (c) One qubit
  per edge, four per block, decouple in state $\lvert + \rangle$. The
  isometry also traces out the twelve decoupled qubits. (d) State of
  the system after the RG transformation.}
\end{figure}

These operations can be inverted to \emph{decouple} qubits in states
$\lvert 0 \rangle$ and $\lvert + \rangle$ from the rest system. The
disentanglers and isometries, defining both the RG transformation and
the MERA for the Kitaev states, are made of several of these
decoupling moves. We regard the original square lattice $\Lambda$, on
which the toric code is defined, as a (tilted) square lattice
${\mathcal{L}}_0$ where each site contains four qubits.  Then both
disentanglers and isometries act on blocks of four sites of
${\mathcal{L}}_0$ as in figure \ref{figure:2DMERA} --- equivalently,
on blocks of 16 qubits in $\Lambda$.  They consist of a series of
CNOTs as specified in figures \ref{figure:meraa} and
\ref{figure:merab}.


Upon applying the RG transformation, we obtain a coarse-grained
lattice ${\cal L}_{1}$ which is locally identical to ${\cal L}_0$ and
where, by construction, the toric code constraints are still
satisfied. This is quite remarkable. On the one hand, it is the first
non-trivial example, in the context of entanglement renormalization,
where the RG transformation is \emph{exact} \cite{exactER}, leading to
the first non-trivial model that can be \emph{exactly} described with
the MERA. On the other hand, if we consider an infinite lattice, the
above observation implies that Kitaev states are an explicit fixed
point of the RG flow in the space of ground states, as induced by the
present RG transformation \cite{fixed_point}.

Let us now consider a finite lattice ${\cal L}_0$ on the torus. The
coarse-grained state carries exactly the same topological information
(values of $\prod Z$ along nontrivial cycles) as the original state,
since the elementary moves preserve such information at each
intermediate step. That is, different Kitaev states are not mixed
during the RG transformation. By iteration, we obtain a sequence of
increasingly coarse-grained lattices $\{{\cal L}_0, {\cal L}_1, {\cal
  L}_2, \cdots, {\cal L}_T \}$ for ever smaller toruses. The top
lattice ${\cal L}_T$ will contain only a few qubits. Recall that the
MERA is made of all the disentanglers and isometries used in the RG
transformations, together with a top tensor describing the state of
${\cal L}_T$ \cite{MERA}. It follows that the MERAs for different
states of the toric code will contain identical disentanglers and
isometries, and will only differ in their top tensor, where all the
topological information is stored.

\begin{figure}
\centering
(a)%
\scalebox{.42}{%
\includegraphics{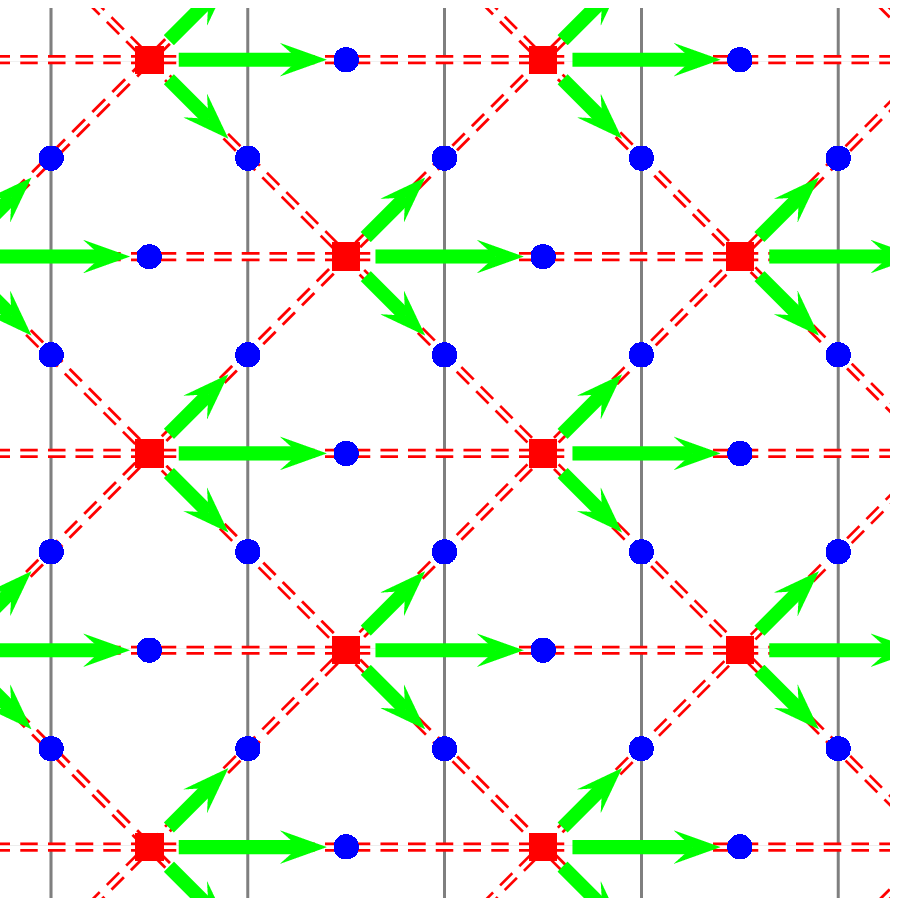}
}%
(b)%
\scalebox{.42}{%
\includegraphics{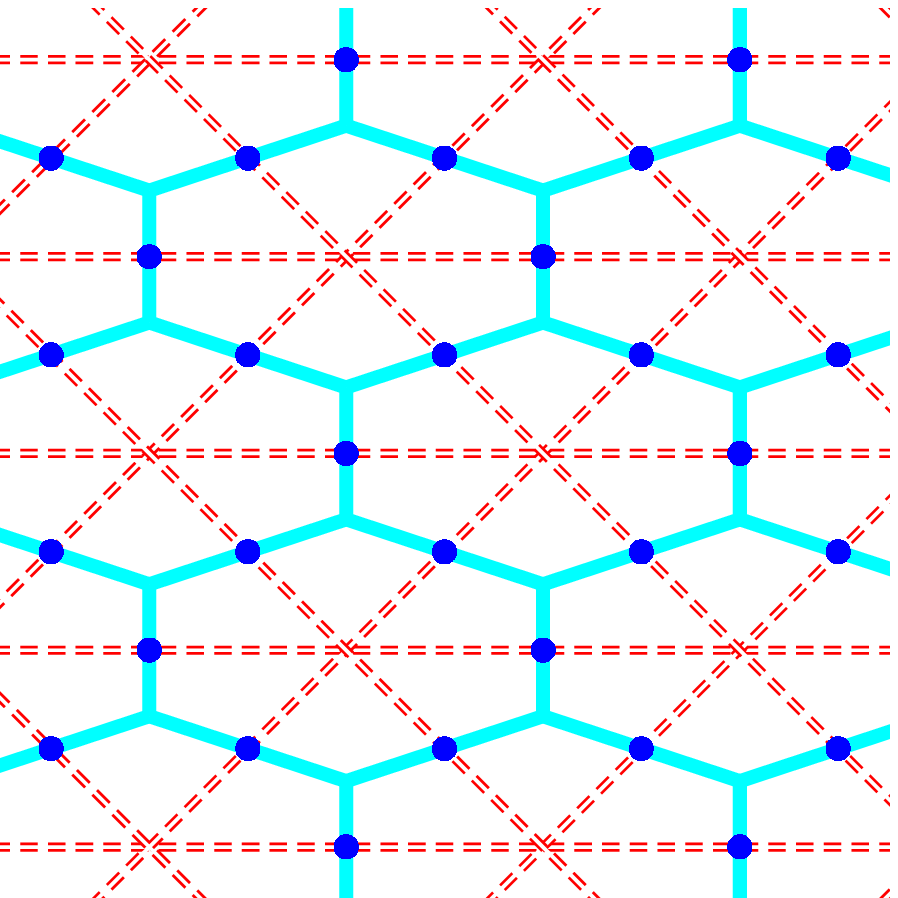}
}
\caption{\label{figure:merabis}%
  Local mapping between the toric code on a square lattice (a) and on
  a triangular lattice (b).  The dual model in a honeycomb lattice
  (displayed for reference) is Levin and Wen's loop model.  }
\end{figure}

All the above results automatically extend to the loop model
considered by Levin and Wen as the simplest of their family of
string-net models \cite{string-net}. Indeed, the toric code on a
square lattice can be locally transformed, using the decoupling moves
depicted in figure \ref{figure:merabis}, into a toric code on a
triangular lattice, which is equivalent to the ground state of the
loop model defined on the dual (hexagonal) lattice. This local
transformation shows that the topological order of both models are
identical, a fact already pointed out in \cite{nussinov} and which can
also be understood in terms of the projected entangled-pair state
ansatz (PEPS) \cite{power}.

Finally, our construction generalizes almost straightfowardly to
quantum double models (see, e.g., \cite{kitaev03}), both for Abelian
and non-Abelian groups.  This is achieved by replacing CNOTs with
controlled group multiplication operators and by paying due attention
to the order of the operations (see appendix).

In conclusion, we have shown that several models with topological
order can be exactly represented with the MERA, where topological
degrees of freedom are naturally isolated in its top tensor. We have
also seen that such models are fixed points of the RG flow induced by
entanglement renormalization.  Our results are an unambiguous sign
that entanglement renormalization and the MERA, originally developed
to efficiently simulate systems with local symmetry-breaking phases,
provide also a most natural framework to study topological phases.

\textbf{Acknowledgements:} We thank J.~I.~Cirac, A.~Kitaev,
D.~P\'erez-Garc\'{\i}a, J.~Preskill and F.~Verstraete for related
discussions.  M.~A.~thanks the University of Queensland for
hospitality and a stimulating working atmosphere during his visit.
G.~V.~acknowledges financial support from the Australian Research
Council, FF0668731.

\begin{appendix}

\section{ Exact MERA for quantum double models}

Here we generalize the above RG transformation and MERA to lattice
quantum double models (see \cite{kitaev03}.)  Local degrees of freedom
are associated with \emph{oriented} bonds of a lattice $\Lambda$ and
identified with the group algebra of a discrete, in general
non-Abelian, group $G$, i.e., the Hilbert space spanned by an
orthonormal basis $\{ \lvert g \rangle, \, g \in G \}$.  A change in
the orientation of a bond corresponds to the map $S : \lvert g \rangle
\mapsto \lvert g^{-1} \rangle$.  The Hamiltonian is a sum of mutually
commuting projectors over vertices and plaquettes,
\begin{equation}\label{qdoublehamiltonian}
  H_{ \mathrm{D} (G) }
=
 - \,
  \sum_v A_v
 - \,
  \sum_p B_p
\end{equation}
where vertex projector $A_v$ acts on edges incoming to vertex $v$ by
simultaneous right multiplication by each group element,
\begin{equation}\label{qdoublevertices}
  A_v
=
  \frac{ 1 }{ \lvert G \rvert } \,
  \sum_{ h \in G }  
  \bigotimes_{i \rightarrow v} R_i ( h )
  \; ,
\end{equation}
right multiplication acts as $R(h) \lvert g \rangle = \lvert g h
\rangle$, and \nobreak{plaquette} projector $B_p$ selects
configurations where the ordered product of group elements taken along
an oriented circuit ${\mathcal{C}}_p$ around $p$ is the unit element
of $G$,
\begin{equation}\label{qdoubleplaquettes}
  B_p
=
  \delta (
    \prod_{ i \text{ along } {\mathcal{C}}_p } g_i, \,
    e
  )
  \; .
\end{equation}

\begin{figure}[h]
\centering
(a)%
\scalebox{.5}{%
\includegraphics{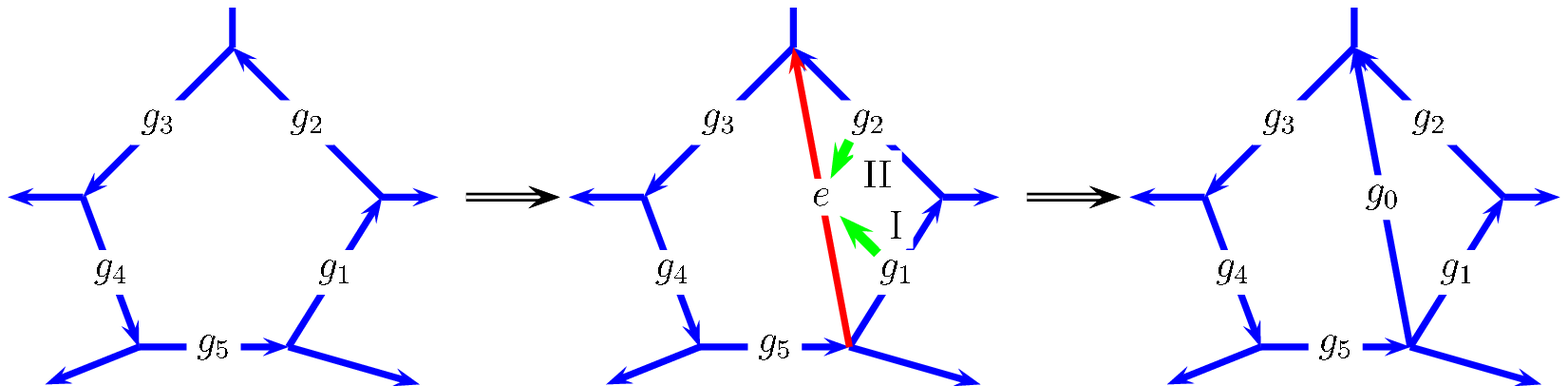}
}

\vspace{1cm}

(b)%
\scalebox{.5}{%
\includegraphics{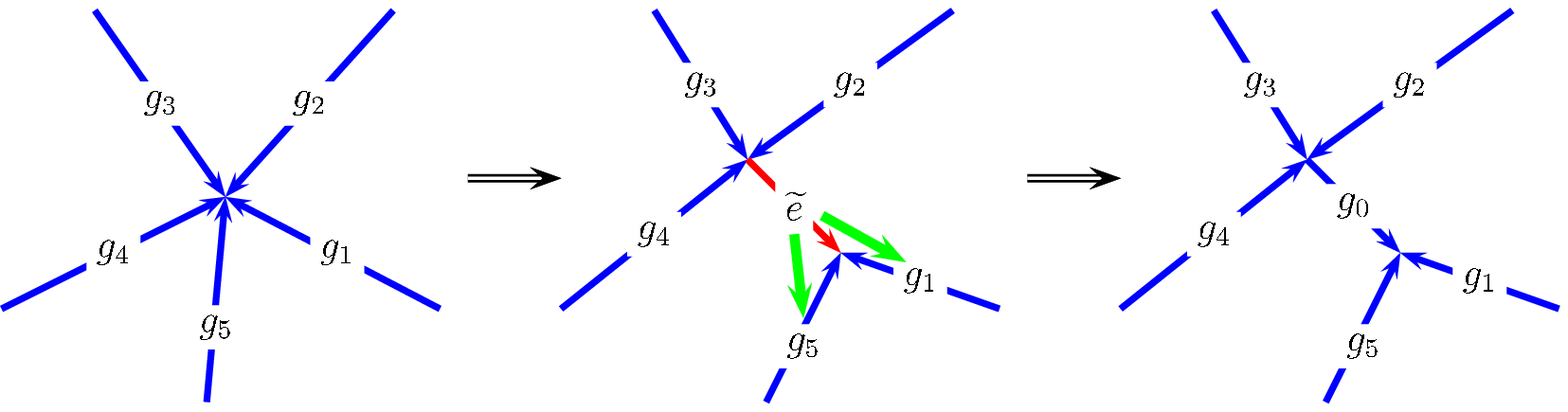}
}
\caption{\label{figure:qdelemmoves}%
  Elementary moves adding plaquettes and vertices to a quantum double
  model.  Local degrees of freedom live in the group algebra of a
  discrete group $G$.  The ancilla is initialised in state $\lvert e
  \rangle$ (unit element of $G$.)  Thick arrows stand for controlled
  right-multiplication of the target element by the control
  element. Orientation of the edges plays an important r\^ole.  (a)
  For plaquette addition, operations must be performed in a prescribed
  order (e.g., after application of the arrows in counterclockwise
  order (I, II), the new element becomes $g_1 g_2$.)  (b) For vertex
  addition, the new element is initialised in state $\lvert
  \widetilde{e} \rangle$, the equal-weight superposition of all
  elements in $G$.  Here all operations can be performed
  simultaneously.}
\end{figure}

Elementary moves are analogous to their counterparts for the toric
code.  The operations generalising CNOTs are controlled
multiplications by the control element (CMs).  Figure
\ref{figure:qdelemmoves} shows how to create plaquettes and vertices
using the controlled right multiplication
\begin{equation}\label{crm}
  A \, \lvert h, \, g \rangle
=
  \lvert h, \, g h \rangle ,
\end{equation}
where the first element is the control and the second element is the
target.  To cover the case of different bond orientations, we also
consider the transformations $B = (S \otimes 1) A (S \otimes 1)$, $C =
(1 \otimes S) A (1 \otimes S)$, and $D = (S \otimes S) A (S \otimes
S)$; explicitly:
\begin{align}\label{othercm}
\nonumber
  B \, \lvert h, \, g \rangle
&=
  \lvert h, \, g h^{-1} \rangle ,
\\
\nonumber
  C \, \lvert h, \, g \rangle
&=
  \lvert h, \, h^{-1} g \rangle ,
\\
  D \, \lvert h, \, g \rangle
&=
  \lvert h, \, h g \rangle .
\end{align}
\begin{figure}[h]
\centering
(a)
\scalebox{.8}{%
%
\includegraphics{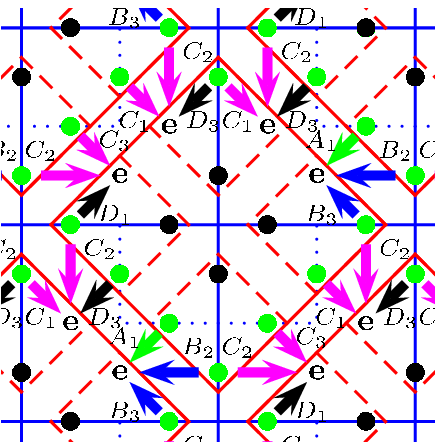}
}%
%
%
(b)
\scalebox{.8}{%
%
\includegraphics{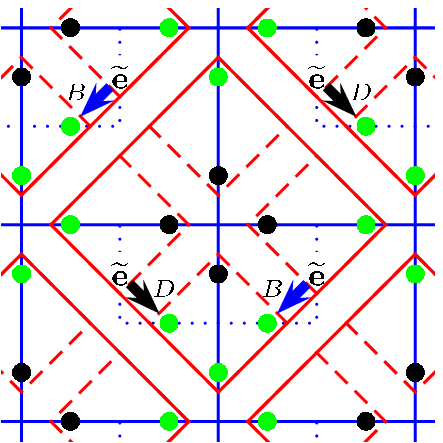}
}

\vspace{.3cm}

(c)
\scalebox{.8}{
%
\includegraphics{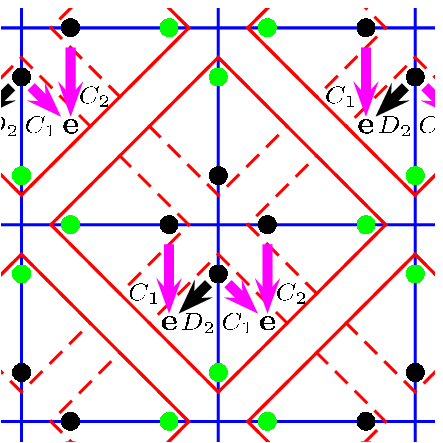}
}%
(d)%
\scalebox{.8}{
%
\includegraphics{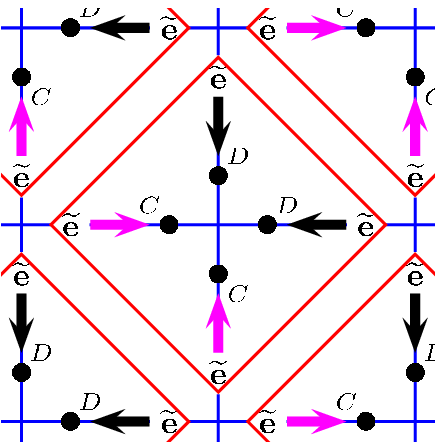}
}
\caption{\label{figure:qdmera}%
  RG transformation for a quantum double model.  Arrows stand for
  controlled multiplications (CM) from control to target elements.
  Arrow labels denote the type of CM (see equations (\ref{crm}) and
  (\ref{othercm})), as well as the order in which they are applied
  within each step.  The fiducial lattice orientation (horizontal
  bonds pointing to the right, vertical bonds pointing upwards) is
  assumed.  (a) Disentanglers.  Four elements per block decouple in
  state $\lvert e \rangle$.  (b)--(c) Isometries.  In (b), two
  elements per block decouple in state $\lvert \widetilde{e} \rangle$.
  In (c), another two elements per block decouple in state $\lvert e
  \rangle$, and the lattice becomes a doubled square lattice with two
  elements per edge.  One element per edge, four per block, then
  decouple in (d) in state $\lvert \widetilde{e} \rangle$, completing
  the MERA ansatz.}
\end{figure}

By means of these operations, new edges initialised in states $\lvert
e \rangle$ and
\begin{equation}\label{etilde}
  \lvert \widetilde{e} \rangle
=
  \frac{ 1 }{ \sqrt{ \lvert G \rvert } }
  \sum_{ h \in G }
  \lvert h \rangle
\end{equation}
are incorporated into the code, creating new plaquettes and vertices.
Of course, the inverse elementary moves \emph{removing} plaquettes and
vertices from the code, needed for the MERA construction, are in
general not identical to those adding plaquettes and vertices.  Note
that operations leading to plaquette addition (or removal) cannot be
performed simultaneously for non-Abelian groups, since the order of
multiplication of the elements is important.

The RG transformation corresponding to a quantum double model
associated with group $G$ and defined on a square lattice proceeds
along the same lines as for the toric code, but there are qualitative
differences.  To fix the setting, we work with a fiducial orientation
of the bonds: horizontal bonds are oriented from left to right and
vertical bonds are oriented upwards.  Then:
\begin{itemize}
\item %
  Operations within a plaquette cannot be performed simultaneously and
  must be applied in a certain order.  Hence, disentanglers must be
  applied in three steps, while isometries demand another step with
  respect to the toric code RG.
\item %
  Which of the controlled operations $A$, $B$, $C$, $D$ is needed at
  each step depends on the bond orientations.
\end{itemize}

The explicit form of the RG leading to a MERA description of the
quantum double model is shown in figure \ref{figure:qdmera}.  The
basic properties of the toric code MERA (bounded causal cone,
topological degrees of freedom at the top of the tensor network, ER
fixed point in the infinite lattice limit) generalise to the quantum
double setting.

\end{appendix}



\begin{thebibliography}{99}
\bibitem{large_dimension} %
  In a $D$ dimensional lattice the ground state typically obeys a
  boundary law $S_l \sim l^{D-1}$ for the entanglement entropy of a
  block of $l^D$ sites. In this case the dimension $d_{\tau}$ for a
  site of ${\mathcal{L}}_\tau$ must at least scale doubly
  exponentially in $\tau$, $d_{\tau} \sim \exp(\exp(\tau))$. Indeed,
  one the one hand the dimension of an effective space for a block of
  $l^D$ sites must be at least $d \sim \exp(S_l) = \exp(l^{D-1})$. On
  the other, $l_{\tau} \sim \exp(\tau)$ after $\tau$ iterations of the
  RG transformation, where $n_{\tau}=l_{\tau}^{D}$ is the number of
  sites of ${\mathcal{L}}_0$ effectively described by a single site of
  ${\mathcal{L}}_\tau$.
\bibitem{fixed_point} %
  This points to a more general result, to be discussed elsewhere:
  under the present RG transformation based on entanglement
  renormalization, systems with topological order flow toward the
  string-net models of Levin and Wen \cite{string-net}, substantiating
  ideas already advocated by these authors.
\end{thebibliography}
\end{document}